\documentclass[preprint,12pt]{elsarticle}

 \usepackage{graphicx}

\usepackage{amssymb}

\journal{Nuclear Physics A}

\begin{document}

\begin{frontmatter}



\title{Average and recommended half-life values for two neutrino double beta decay}


\author{A.S.~Barabash} 

\address{Institute of Theoretical and Experimental Physics, B. Cheremushkinskaya 25, 
117218 Moscow, Russia}

\begin{abstract}

All existing positive results on two neutrino double beta decay in 
different nuclei were analyzed.  Using the procedure recommended by the 
Particle Data Group, weighted average values for half-lives of 
$^{48}$Ca, $^{76}$Ge, $^{82}$Se, $^{96}$Zr, $^{100}$Mo, $^{100}$Mo - 
$^{100}$Ru ($0^+_1$), $^{116}$Cd, $^{130}$Te, $^{136}$Xe, $^{150}$Nd, $^{150}$Nd - $^{150}$Sm 
($0^+_1$) and $^{238}$U were obtained. Existing geochemical data were 
analyzed and recommended values for half-lives of $^{128}$Te, 
and $^{130}$Ba are proposed. Given the measured half-life values, nuclear matrix elements 
were calculated using latest (more reliable and precise) values for phase space factor.
Finally, previous results (PRC 81 (2010) 035501) were up-dated and results for $^{136}$Xe 
were added.
  
\end{abstract}


\begin{keyword}
Double beta decay, Nuclear matrix elements, $^{48}$Ca, $^{76}$Ge, $^{82}$Se, $^{96}$Zr, $^{100}$Mo, 
$^{100}$Mo - 
$^{100}$Ru ($0^+_1$), $^{116}$Cd, $^{128}$Te, $^{130}$Te, $^{136}$Xe, $^{150}$Nd, $^{150}$Nd - $^{150}$Sm 
($0^+_1$), $^{238}$U and $^{130}$Ba

\end{keyword}

\end{frontmatter}


\section{Introduction}
\label{}

At present, the two neutrino double beta ($2\nu\beta\beta$) decay process 
has been detected in a total of 11 different nuclei. In $^{100}$Mo and 
$^{150}$Nd, this type of decay was also detected for the transition to
the $0^+$ excited state of the daughter nucleus. For the case of the 
$^{130}$Ba nucleus, evidence for the two neutrino double electron capture 
process was observed via a geochemical experiments.  All of these results 
were obtained in a few tens of geochemical experiments and more than forty 
direct (counting) experiments as well as and in one radiochemical experiment. In 
direct experiments, for some nuclei there are as many as eight independent 
positive results (e.g., $^{100}$Mo).  In some experiments, the statistical 
error does not always play the primary role in overall half-life 
uncertainties. For example, the NEMO-3 experiment with $^{100}$Mo has currently detected 
more than 219,000 $2\nu\beta\beta$ events \cite{ARN05}, which results in a value for 
the statistical error of $\sim$ 0.2\% . At the same time, the systematic 
error for many experiments on $2\nu\beta\beta$ decay  
remains quite high ($\sim 10-30\%$) and very often cannot be determined 
reliably.  As a consequence, it is frequently difficult for the 
``user'' to select the ``best'' half-life value among the 
results. Using an averaging procedure, one can produce the most 
reliable and accurate half-life values for each isotope. 

Why are accurate 
half-life periods necessary? The most important motivations 
are the following:\\
{\it 1) Nuclear spectroscopy}. Now we know that some isotopes which were earlier 
considered to be stable are not, and decay via the double beta decay processes 
with a half-life period of
 $\sim 10^{18}-10^{24}$ yr are observed. The values which are presented here should be 
introduced into the isotope tables.\\
{\it 2) Nuclear matrix elements (NME)}. First, it gives the possibility to improve 
the quality of NME calculations for two neutrino double beta decay, so 
one can directly compare experimental and calculated values. 
For example, so-called "$g_A$ (axial-vector coupling constant) quenching" problem 
could be solved by comparison of exact experimental values of NMEs and results of 
theoretical calculations (see discussions in Ref. \cite{FAE08,CAU12,SUH13,BAR13a}. 
Second, 
it gives the possibility to improve 
the quality of NME calculations for neutrinoless double beta decay. 
The accurate half-life values for $2\nu\beta\beta$ decay
are used to adjust the most relevant parameter of the 
quasiparticle random-phase approximation (QRPA) model, the strength of the 
particle-particle interaction $g_{pp}$ \cite{ROD06,KOR07,KOR07a,SIM08}.\\
{\it 3) Research on the single state dominance (SSD) mechanism \cite{SIM01,DOM05} and a check of 
the "bosonic" component of the neutrino hypothesis \cite{DOL05,BAR07} is possible}.

In the present work, an analysis of all ``positive'' experimental
results has been performed, and averaged or recommended values for  
isotopes are presented.

The first time that this work was done was in 2001, and the 
results were presented at the International Workshop on the calculation of double beta decay 
nuclear matrix elements, MEDEX'01 \cite{BAR02}. Then revised half-life values were
presented at MEDEX'05 and MEDEX'09 and published in 
Ref. \cite{BAR06} and \cite{BAR09a,BAR10}, respectively.  
In the present paper, new positive results obtained since end of 2009 
and to the end of 2014 have been added and analyzed.

The main differences from the previous analysis \cite{BAR10} are the following:

1) The new experimental data obtained after the publication of Ref. \cite{BAR10} 
are included in the analysis:  $^{76}$Ge \cite{AGO13}, $^{100}$Mo \cite{CAR14}, 
$^{100}$Mo - $^{100}$Ru($0^+_1$) \cite{BEL10,ARN14}, $^{116}$Cd \cite{POD14}, 
$^{130}$Te \cite{ARN11}, $^{150}$Nd - $^{150}$Sm (0$^+_1$) \cite{KID14} and $^{130}$Ba\cite{PUJ09}.

2) "Positive" results obtained for $^{136}$Xe \cite{GAN12,ALB14} are analyzed. 
This decay was detected for the fist time in 2011 \cite{ACK11}.

3) To calculate NMEs new phase space factor values ($G_{2\nu}$) from Ref. \cite{KOT12,KOT13} 
and \cite{STO13,STO14} are used.

4) Considering possible changes of axial vector coupling constant $g_A$ 
(possible quenching effect in nuclear medium) so-called "effective" NMEs are calculated,  
$\mid M^{eff}_{2\nu}\mid = g_A^2\cdot \mid (m_ec^2\cdot M_{2\nu})\mid$ 
(in Ref. \cite{BAR10} the dimensionless nuclear matrix elements $\mid (m_ec^2\cdot M_{2\nu})\mid$  
were calculated for $g_A$ = 1.254).

\section{Present experimental data}

Experimental results on $2\nu\beta\beta$ decay in different nuclei are 
presented in Table 1.  For direct experiments, the number of events 
and the signal-to-background ratio are presented.

\begin{table}
\caption{Present, positive $2\nu\beta\beta$ decay results. 
Here N is the number of useful events, $T_{1/2}$ is a half-life and S/B is the signal-to-background 
ratio.
\newline
$^{a)}$ For $E_{2e} > 1.2$ MeV; $^{b)}$ after correction (see text); 
$^{c)}$ for the SSD mechanism; $^{d)}$ for $E_{2e} > 1.5$ MeV;  $^{e)}$ in both peaks.}
\begin{tabular}{|c|c|c|c|c|}
\hline
\rule[-2.5mm]{0mm}{6.5mm}
Nucleus & N & $T_{1/2}$, yr & S/B & Ref., year \\
\hline
\rule[-2mm]{0mm}{6mm}
$^{48}$Ca & $\sim 100$ & $[4.3^{+2.4}_{-1.1}(stat)\pm 1.4(syst)]\cdot 10^{19}$
  & 1/5 & \cite{BAL96}, 1996 \\
 & 5 & $4.2^{+3.3}_{-1.3}\cdot 10^{19}$ & 5/0 & \cite{BRU00}, 2000 \\
& 116 & $[4.4^{+0.5}_{-0.4}(stat)\pm 0.4(syst)\cdot 10^{19}$ & 6.8 & \cite{BAR11a}, 2011 \\
\rule[-4mm]{0mm}{10mm}
 & & {\bf Average value:} $\bf 4.4^{+0.6}_{-0.5} \cdot 10^{19}$ & & \\  
          
\hline
\rule[-2mm]{0mm}{6mm}
$^{76}$Ge & $\sim 4000$ & $(0.9\pm 0.1)\cdot 10^{21}$ & $\sim 1/8$                                                        
& \cite{VAS90}, 1990 \\
& 758 & $1.1^{+0.6}_{-0.3}\cdot 10^{21}$ & $\sim 1/6$ & \cite{MIL91}, 1991 \\
& $\sim 330$ & $0.92^{+0.07}_{-0.04}\cdot 10^{21}$ & $\sim 1.2$ & \cite{AVI91}, 1991 \\
& 132 & $1.27^{+0.21}_{-0.16}\cdot 10^{21}$ & $\sim 1.4$ & \cite{AVI94}, 1994 \\
& $\sim 3000$ & $(1.45\pm 0.15)\cdot 10^{21}$ & $\sim 1.5$ & \cite{MOR99}, 1999 
\\
& $\sim 80000$ & $[1.74\pm 0.01(stat)^{+0.18}_{-0.16}(syst)]\cdot 10^{21}$ & $\sim 1.5$ 
& \cite{HM03}, 2003 \\
& 7030 & $1.84^{+0.14}_{-0.10}\cdot 10^{21}$ & $\sim 4$ & \cite{AGO13}, 2013 \\
\rule[-4mm]{0mm}{10mm}
& & {\bf Average value:} $\bf 1.65^{+0.14}_{-0.12} \cdot 10^{21}$ & & \\

\hline
\rule[-2mm]{0mm}{6mm}

$^{82}$Se& 89.6 & $1.08^{+0.26}_{-0.06}\cdot 10^{20}$ & $\sim 8$ & \cite{ELL92}, 1992 \\
 & 149.1 & $[0.83 \pm 0.10(stat) \pm 0.07(syst)]\cdot 10^{20}$ & 2.3 & 
\cite{ARN98}, 1998 \\
& 2750 & $[0.96 \pm 0.03(stat) \pm 0.1(syst)]\cdot 10^{20}$ & 4 & \cite{ARN05}, 2005\\ 
& & $(1.3\pm 0.05)\cdot 10^{20}$ (geochem.) & & \cite{KIR86}, 1986 \\
\rule[-4mm]{0mm}{10mm}
& & {\bf Average value:} $\bf (0.92\pm 0.07)\cdot 10^{20}$ & & \\
 
\hline
\rule[-2mm]{0mm}{6mm}
$^{96}$Zr & 26.7 & $[2.1^{+0.8}_{-0.4}(stat) \pm 0.2(syst)]\cdot 10^{19}$ & $1.9^{a)}$ 
& \cite{ARN99}, 1999 \\
& 453 & $[2.35 \pm 0.14(stat) \pm 0.16(syst)]\cdot 10^{19}$ & 1 & \cite{ARG10}, 2010\\
& & $(3.9\pm 0.9)\cdot 10^{19}$ (geochem.)& & \cite{KAW93}, 1993 \\
& & $(0.94\pm 0.32)\cdot 10^{19}$ (geochem.)& & \cite{WIE01}, 2001 \\
\rule[-4mm]{0mm}{10mm}
& & {\bf Average value:} $\bf (2.3 \pm 0.2)\cdot 10^{19}$ & & \\

\hline

\end{tabular}
\end{table}

\addtocounter{table}{-1}
\begin{table}
\caption{continued.}
\bigskip
\begin{tabular}{|c|c|c|c|c|}

\hline
\rule[-2mm]{0mm}{6mm}
$^{100}$Mo & $\sim 500$ & $11.5^{+3.0}_{-2.0}\cdot 10^{18}$ & 1/7 & 
\cite{EJI91}, 1991 \\
& 67 & $11.6^{+3.4}_{-0.8}\cdot 10^{18}$ & 7 & \cite{ELL91}, 1991 \\
& 1433 & $[7.3 \pm 0.35(stat) \pm 0.8(syst)]\cdot 10^{18b)}$ & 3 & 
\cite{DAS95}, 1995 \\
& 175 & $7.6^{+2.2}_{-1.4}\cdot 10^{18}$ & 1/2 & \cite{ALS97}, 1997 \\
& 377 & $[6.75^{+0.37}_{-0.42}(stat) \pm 0.68(syst)]\cdot 10^{18}$ & 10 & 
\cite{DES97}, 1997 \\
& 800 & $[7.2 \pm 1.1(stat) \pm 1.8(syst)]\cdot 10^{18}$ & 1/9 & 
\cite{ASH01}, 2001 \\
& 219000 & $[7.11 \pm 0.02(stat) \pm 0.54(syst)]\cdot 10^{18c)}$ & 40 & 
\cite{ARN05}, 2005\\
& $\sim$ 350 & $[7.15 \pm 0.37(stat) \pm 0.66(syst)]\cdot 10^{18}$ & $\sim$ $5^{d)}$ & 
\cite{CAR14}, 2014\\
& & $(2.1\pm 0.3)\cdot 10^{18}$ (geochem.)& & \cite{HID04}, 2004 \\ 
\rule[-4mm]{0mm}{10mm}
& & {\bf Average value:} $\bf (7.1\pm 0.4)\cdot 10^{18}$ & & \\

\hline
$^{100}$Mo - & $133^{e)}$ & $6.1^{+1.8}_{-1.1}\cdot 10^{20}$ & 1/7 & 
\cite{BAR95}, 1995 \\
$^{100}$Ru ($0^+_1$) &  $153^{e)}$ & $[9.3^{+2.8}_{-1.7}(stat) \pm 1.4(syst)]\cdot 
10^{20}$ & 1/4 & \cite{BAR99}, 1999 \\
 & 19.5 & $[5.9^{+1.7}_{-1.1}(stat) \pm 0.6(syst)]\cdot 10^{20}$ & $\sim 8$ & 
\cite{DEB01}, 2001 \\ 
& 35.5 & $[5.5^{+1.2}_{-0.8}(stat) \pm 0.3(syst)]\cdot 10^{20}$ & $\sim 8$ & 
\cite{KID09}, 2009 \\ 
& 37.5 & $[5.7^{+1.3}_{-0.9}(stat) \pm 0.8(syst)]\cdot 10^{20}$ & $\sim 3$ & 
\cite{ARN07}, 2007 \\ 
& $597^{e)}$ & $[6.9^{+1.0}_{-0.8}(stat) \pm 0.7(syst)]\cdot 10^{20}$ & $\sim 1/10$ & 
\cite{BEL10}, 2010 \\
& $239^{e)}$ & $[7.5 \pm 0.6(stat) \pm 0.6(syst)]\cdot 10^{20}$ & 2 & 
\cite{ARN14}, 2014 \\     
\rule[-4mm]{0mm}{10mm}
& & {\bf Average value:} $\bf 6.7^{+0.5}_{-0.4}\cdot 10^{20}$ & & \\

\hline
$^{116}$Cd& $\sim 180$ & $2.6^{+0.9}_{-0.5}\cdot 10^{19}$ & $\sim 1/4$ & 
\cite{EJI95}, 1995 \\
& 174.6 & $[2.9 \pm 0.3(stat) \pm 0.2(syst)]\cdot 10^{19 b)}$ & 3 & 
\cite{ARN96}, 1996 \\
& 9850 & $[2.9\pm 0.06(stat)^{+0.4}_{-0.3}(syst)]\cdot 10^{19}$ & $\sim 3$ & 
\cite{DAN03}, 2003 \\
& 7000 & $[2.88 \pm 0.04(stat) \pm 0.16(syst)]\cdot 10^{19 c)}$ & 10 & \cite{BAR11}, 2011\\
& 34927 & $[2.80 \pm 0.05(stat) \pm 0.4(syst)]\cdot 10^{19}$ & 2 & \cite{POD14}, 2014\\
\rule[-4mm]{0mm}{10mm}
& & {\bf Average value:} $\bf (2.87 \pm 0.13)\cdot 10^{19}$ & & \\

\hline
\rule[-2mm]{0mm}{6mm}
$^{128}$Te& & $\sim 2.2\cdot 10^{24}$ (geochem.) & & \cite{MAN91}, 1991 \\
& & $(7.7\pm 0.4)\cdot 10^{24}$ (geochem.)& & \cite{BER93}, 1993 \\
& & $(2.41\pm 0.39)\cdot 10^{24}$ (geochem.)& & \cite{MES08}, 2008 \\
& & $(2.3\pm 0.3)\cdot 10^{24}$ (geochem.)& & \cite{THO08}, 2008 \\
\rule[-4mm]{0mm}{10mm}
& & {\bf Recommended value:} $\bf (2.0\pm 0.3)\cdot 10^{24}$ & & \\

\hline
\end{tabular}
\end{table}

\addtocounter{table}{-1}
\begin{table}
\caption{continued 2.}
\bigskip
\begin{tabular}{|c|c|c|c|c|}

\hline
\rule[-2mm]{0mm}{6mm}
$^{130}$Te& 260 & $[6.1 \pm 1.4(stat)^{+2.9}_{-3.5}(syst)]\cdot 10^{20}$ & 1/8 & \cite{ARN03}, 2003 \\
& 236 & $[7.0 \pm 0.9(stat) \pm 1.1(syst)]\cdot 10^{20}$ & 1/3 & \cite{ARN11}, 2011 \\
& & $\sim 8\cdot 10^{20}$ (geochem.) & & \cite{MAN91}, 1991 \\
& & $(27\pm 1)\cdot 10^{20}$ (geochem.)& & \cite{BER93}, 1993 \\
& & $(9.0\pm 1.4)\cdot 10^{20}$ (geochem.)& & \cite{MES08}, 2008 \\
& & $(8.0\pm 1.1)\cdot 10^{20}$ (geochem.)& & \cite{THO08}, 2008 \\
\rule[-4mm]{0mm}{10mm}
& & {\bf Average value:} $\bf (6.9 \pm 1.3)\cdot 10^{20}$ & & \\

\hline
\rule[-2mm]{0mm}{6mm}
$^{136}$Xe & $\sim$ 50000 & $[2.30 \pm 0.02(stat) \pm 0.12(syst)]\cdot 10^{21}$ & 
$\sim$10 & \cite{GAN12}, 2012 \\
&  19042 & $[2.165 \pm 0.016(stat) \pm 0.059(syst)]\cdot 10^{21}$ & $\sim$10 & 
\cite{ALB14}, 2014 \\

\rule[-4mm]{0mm}{10mm}
& & {\bf Average value:} $\bf(2.19\pm 0.06)\cdot 10^{21}$ & & \\

\hline
\rule[-2mm]{0mm}{6mm}
$^{150}$Nd& 23 & $[18.8^{+6.9}_{-3.9}(stat) \pm 1.9(syst)]\cdot 10^{18}$ & 
1.8 & \cite{ART95}, 1995 \\
& 414 & $[6.75^{+0.37}_{-0.42}(stat) \pm 0.68(syst)]\cdot 10^{18}$ & 6 & 
\cite{DES97}, 1997 \\
& 2018 & $[9.11^{+0.25}_{-0.22}(stat) \pm 0.63(syst)]\cdot 10^{18}$ & 2.8 & \cite{ARG08}, 2008\\
\rule[-4mm]{0mm}{10mm}
& & {\bf Average value:} $\bf(8.2\pm 0.9)\cdot 10^{18}$ & & \\

\hline
\rule[-2mm]{0mm}{6mm}
$^{150}$Nd - & $177.5^{d)}$ & $[1.33^{+0.36}_{-0.23}(stat)^{+0.27}_{-0.13}(syst)]\cdot 10^{20}$ & 
1/5 & \cite{BAR09}, 2009 \\
& 21.6 & $[1.07^{+0.45}_{-0.25}(stat) \pm {+0.07}(syst)]\cdot 10^{20}$ & $\sim$ 1.2 & \cite{KID14}, 2014\\
$^{150}$Sm ($0^+_1$) & & {\bf Average value:} $\bf 1.2^{+0.3}_{-0.2}\cdot 10^{20}$ & \\ 
 
\hline
\rule[-2mm]{0mm}{6mm}
$^{238}$U& & $\bf (2.0 \pm 0.6)\cdot 10^{21}$ (radiochem.) &  & \cite{TUR91}, 1991 \\
 
\hline
\rule[-2mm]{0mm}{6mm}
$^{130}$Ba &  & $\bf 2.1^{+3.0}_{-0.8} \cdot 10^{21}$ (geochem.) & 
 & \cite{BAR96}, 1996 \\
 ECEC(2$\nu$)&  & $\bf (2.2 \pm 0.5)\cdot 10^{21}$ (geochem.) & 
 & \cite{MES01}, 2001 \\
& & $\bf (0.60 \pm 0.11)\cdot 10^{21}$ (geochem.) &
& \cite{PUJ09}, 2009 \\
\rule[-4mm]{0mm}{10mm}
& & {\bf Recommended value:} $ \bf \sim 10^{21}$ & & \\
\hline

\end{tabular}
\end{table}

\section{Data analysis}

To obtain an average of the ensemble of available data, a standard weighted 
least squares procedure, as recommended by the Particle Data Group 
\cite{BER12}, was used.  The weighted average and the corresponding error 
were calculated, as follows:
\begin{equation}
\bar x\pm \delta \bar x = \sum w_ix_i/\sum w_i \pm (\sum w_i)^{-1/2} , 
\end{equation} 
where $w_i = 1/(\delta x_i)^2$.  Here, $x_i$ and $\delta x_i$ are 
the value and error reported by the i-th experiment, and 
the summations run over the N experiments.  

The next step is to calculate $\chi^2 = \sum w_i(\bar x - x_i)^2$ and 
compare it with N - 1, which is the expectation value of $\chi^2$ if the 
measurements are from a Gaussian distribution.  If $\chi^2/(N - 1)$ is 
less than or equal to 1 and there are no known problems with the data, 
then one accepts the results to be sound.  If $\chi^2/(N - 1)$ is very large ($>> 1$), 
one chooses 
not to use the average. Alternatively, one may quote the calculated 
average, while making an educated guess of the error, using a conservative 
estimate designed to take into account known problems with the data.
Finally, if $\chi^2/(N - 1)$ is larger than 1, but not greatly so, it is  
still best to use the average data, but to increase the quoted error, $\delta \bar x$ 
in Equation 1, by a factor of S defined by 
\begin{equation}
S = [\chi^2/(N - 1)]^{1/2}.
\end{equation} 
For averages, the statistical and systematic errors are treated in quadrature 
and used as a  combined error $\delta x_i$. In some cases, only the results 
obtained with high enough 
signal-to-background ratio were used. 


\subsection{$^{48}$Ca }    
There are three independent experiments in 
which $2\nu\beta\beta$ decay of $^{48}$Ca was observed \cite{BAL96,BRU00,BAR11a}. 
The results are in good agreement. The weighted average value is:
$$
T_{1/2} = 4.4^{+0.6}_{-0.5} \cdot 10^{19} \rm{yr}.
$$ 

\subsection{$^{76}$Ge } 
Considering the results of six 
experiments, a few additional comments are 
necessary, as follows:

1) We use here final result of the Heidelberg-Moscow Collaboration, 
 $T_{1/2} = [1.74\pm 0.01(stat)^{+0.18}_{-0.16}(syst)]\cdot 10^{21}$ yr
 \cite{HM03}. 

2) In Ref. \cite{AVI91}, the value $T_{1/2} = 
0.92^{+0.07}_{-0.04}\cdot 10^{21}$ yr was presented. However, after a more 
careful analysis, this result has been changed \cite{AVI94}. In Ref. \cite{AVI94} 
a few values for half-life using different analysis methods were obtained. 
I use here the value obtained by fit the data using $\chi^2$ model, which 
take into account shape of the spectrum (see Table 1). Unfortunately systematic 
error was not discussed and taken into account in this paper. This is why during 
my analysis I added typical systematic error for such sort of experiments  ($\pm 10\%$). 
So, finally, I use $T_{1/2} = [1.27^{+021}_{-0.16}(stat) \pm 0.13(syst)] \cdot10^{21}$ yr 
as a result of Ref.  \cite{AVI94}. 

3) The results presented in Ref. \cite{VAS90} do not agree with the more 
recent experiments \cite{MOR99,HM03,AGO13}.   Furthermore, the 
error presented in \cite{VAS90} appears to be too small, especially taking 
into account that the signal-to-background ratio in this 
experiment is equal to $\sim 1/8$. It has been mentioned before
\cite{BAR90} that the half-life value in this work can be $\sim 1.5-2$ 
times higher because the thickness of the dead layer in the Ge(Li) 
detectors used can be different for crystals made from enriched Ge, rather 
than natural Ge. With no uniformity of the external background 
(and this is the case!), this 
effect can have an appreciable influence on the final result.

Finally, in calculating the average, only the results of experiments 
with signal-to-background ratios greater than 1 were used (i.e., the 
results of Refs. \cite{AVI94,MOR99,HM03,AGO13}). The weighted average value is:
$$
    T_{1/2} = 1.65^{+0.14}_{-0.12} \cdot 10^{21} \rm{yr}.
$$ 

\subsection{$^{82}$Se}
There are three independent counting 
experiments and many geochemical measurements $(\sim 20)$ for $^{82}$Se. The geochemical 
data are neither in good agreement with each other nor in good agreement 
with the data from the direct measurements. Typically, the accuracy of 
geochemical measurements is at the level of 10\% and
sometimes even better.  Nevertheless, the possibility of existing large 
systematic errors cannot be excluded (see discussion in Ref. \cite{MAN86}). 
 Thus, to obtain a 
present half-life value for $^{82}$Se, only the results of the direct 
measurements \cite{ARN05,ELL92,ARN98} were used.  The result of Ref. 
\cite{ELL87} is the preliminary result of \cite{ELL92}; hence it has not
been used in our analysis. The result of work \cite{ELL92} is presented with very 
asymmetrical errors. To be more conservative only the top 
error in this case is used. As a result, the weighted average value is:
$$
T_{1/2} = (0.92 \pm 0.07) \cdot 10^{20} \rm{yr}.
$$ 

\subsection{$^{96}$Zr} 
There are two positive geochemical results
\cite{KAW93,WIE01} and two results from the direct experiments of NEMO-2 \cite{ARN99} and 
NEMO-3 \cite{ARG10}.  Taking into account the comment in 
Sec. 3.3, I use the values from Refs. \cite{ARN99,ARG10} to obtain 
a present weighted half-life value for $^{96}$Zr of: 
$$
T_{1/2} = (2.3 \pm 0.2)\cdot 10^{19} \rm{yr}.                    
$$ 

\subsection{$^{100}$Mo}
 
There are eight positive 
results from direct experiments\footnote{I do not consider the result 
of Ref. \cite {VAS90a} because of a potentially high background contribution 
that was 
not excluded in this experiment.} and one 
result from a geochemical experiment. I do not consider the 
preliminary result of Elliott et al. \cite{ELL91} and instead use their 
final result \cite{DES97}, plus I do not use the geochemical result 
(again, see comment in Sec. 3.3).  Finally, in calculating the average, 
only the results of experiments with signal-to-background
ratios greater than 1 were used (i.e., the results of Refs. 
\cite{DAS95,DES97,ARN05,CAR14}).  In addition, I have used the corrected 
half-life value from Ref. \cite {DAS95} (see explanation in \cite{BAR10}).  
The following weighted average value 
for this half-life is then obtained as:
$$
T_{1/2} = (7.1 \pm 0.4)\cdot 10^{18} \rm{yr}.                                   
$$

\subsection{$^{100}$Mo - $^{100}$Ru ($0^+_1$; 1130.32 keV)} 

The 
transition to the $0^+_1$ excited state of $^{100}$Ru was detected in seven 
independent experiments.  The results are in good agreement, and the 
weighted average for the half-life using the results from \cite{BAR95,BAR99,KID09,ARN07,BEL10,ARN14} is:
$$
T_{1/2} = 6.7^{+0.5}_{-0.4} \cdot 10^{20} \rm{yr} .
$$                                   
The result from \cite{DEB01} was not used here because we considered the result from \cite{KID09}
as the final result of the TUNL-ITEP experiment.

\subsection{$^{116}$Cd}
 
There are five independent positive 
results \cite{BAR11a,EJI95,DAN03,ARN96,POD14} that are in good agreement with each other when taking into 
account the corresponding error bars.  Again, I use here the corrected 
result for the half-life value from Ref. \cite{ARN96}.  The original 
half-life value was decreased by $\sim$ 25\% (see explanation in \cite{BAR10}). The 
weighted average value is: 
$$          
T_{1/2} = (2.87 \pm 0.13)\cdot 10^{19} \rm{yr}.
$$ 

\subsection{$^{128}$Te and $^{130}$Te} 

For a long time, there were only geochemical 
data for these isotopes. Although the half-life 
ratio for these isotopes has been obtained with good accuracy $(\sim 3\%)$ 
\cite{BER93}, the absolute values for $T_{1/2}$ of each nuclei 
are different from one experiment to the next.  One group of authors 
\cite{MAN91,TAK66,TAK96} gives $T_{1/2} \approx 0.8\cdot 10^{21}$ yr  
for $^{130}$Te and $T_{1/2} \approx  2\cdot 10^{24}$ yr for $^{128}$Te, 
whereas another group \cite{KIR86,BER93} claims $T_{1/2} \approx 
(2.5-2.7)\cdot 10^{21}$ yr and  $T_{1/2} \approx 7.7\cdot 10^{24}$ yr, 
respectively. Furthermore, as a rule, experiments with young 
samples ($\sim 100$ million years) give results of the half-life value of 
$^{130}$Te in the range of $\sim (0.7-0.9)\cdot 10^{21}$ yr,
while old samples ($> 1$ billion years) have half-life values in the
range of $\sim (2.5-2.7)\cdot 10^{21}$ yr. 
Recently it was argued that short half-lives are more likely to be correct \cite{MES08,THO08}.
Using different young mineral results, the half-life values were estimated at 
$(9.0 \pm 1.4)\cdot 10^{20}$ yr \cite{MES08} and $(8.0 \pm 1.1)\cdot 10^{20}$ yr \cite{THO08} 
for $^{130}$Te and $(2.41 \pm 0.39)\cdot 10^{24}$ y \cite{MES08} and $(2.3 \pm 0.3)\cdot 10^{24}$ yr \cite{THO08} 
for $^{128}$Te. 

The first indication of a positive result for $^{130}$Te in a direct experiment was obtained in 
\cite{ARN03}. More accurate and reliable value was obtained recently in NEMO-3 experiment \cite{ARN11}.
The results are in good agreement, and the weighted average value for half-life is
$$
T_{1/2} = (6.9 \pm 1.3)\cdot 10^{20} yr.
$$ 
Now, using very well-known ratio $T_{1/2}(^{130}{\rm Te})/T_{1/2}(^{128}{\rm Te}) =
(3.52 \pm 0.11)\cdot 10^{-4}$ \cite{BER93},
one can obtain half-life value for $^{128}$Te,
$$
T_{1/2} = (2.0 \pm 0.3)\cdot 10^{24} yr.
$$  
I recommend to use these last two results as the best present half-life values for 
$^{130}$Te and $^{128}$Te, respectively.

\subsection{$^{136}$Xe}
The half-life value was recently measured in two independent experiments, EXO \cite{ACK11,ALB14} 
and Kamland-Zen \cite{GAN12a,GAN12}.
To obtain average value I use most precise results from these experiments, obtained in 
\cite{GAN12,ALB14} (see Table 1). 
The weighted average value is
$$
T_{1/2} = (2.19 \pm 0.06)\cdot 10^{21} yr.
$$

\subsection{$^{150}$Nd}

This half-life value was measured in three 
independent experiments \cite{ART95,DES97,ARG08}. The most accurate value was obtained in Ref. 
\cite{ARG08}. This value is higher than in Ref. \cite{DES97} and lower than in Ref. \cite{ART95} 
($\sim 3\sigma$ and $\sim 2\sigma$ differences, respectively). Using Eq. (1), and three 
existing values, one 
obtains $T_{1/2} = (8.2 \pm 0.5)\cdot 10^{18}$ yr.  Taking into account 
the fact that $\chi^2 > 1$ and S = 1.89 (see Eq. (2)) we then obtain:
$$
T_{1/2} = (8.2 \pm 0.9)\cdot 10^{18} \rm{yr}.
$$ 

\subsection{$^{150}$Nd - $^{150}$Sm ($0^+_1$; 740.4 keV)}

There are two independent experiments in 
which $2\nu\beta\beta$ decay of $^{150}$Nd to the $0^+_1$ excited state of $^{150}$Sm 
was observed \cite{BAR09,KID14} (the preliminary result of Ref. \cite{BAR09} was 
published in Ref. \cite{BAR04}). 
The results are in good agreement. The weighted average value is:

$$          
T_{1/2} = 1.2^{+0.3}_{-0.2}\cdot 10^{20} \rm{yr}.
$$

\subsection{$^{238}$U}  
There is only one positive result but this time from 
a radiochemical experiment \cite{TUR91}:
$$          
T_{1/2} = (2.0 \pm 0.6)\cdot 10^{21} \rm{yr}.
$$ 

\begin{table}[ht]
\caption{Half-life and nuclear matrix element values for two neutrino double beta decay 
(see Sec. 4). For $^{130}$Ba $G_{2\nu}$ value 
for ECEC transition is taken from \cite{KOT13}. $^{a)}$ Obtained using SSD model.}
\begin{tabular}{ccccc}
\hline
Isotope & $T_{1/2}(2\nu)$, yr &  
& $\mid M^{eff}_{2\nu}\mid$   &   \\
& & ($G_{2\nu}$ from \cite{KOT12}) & ($G_{2\nu}$ from \cite{STO14}) & recommended\\
& & & & value\\
\hline
$^{48}$Ca & $4.4^{+0.6}_{-0.5}\cdot10^{19}$ & $0.0382^{+0.0024}_{-0.0024}$ 
& $0.0382^{+0.0024}_{-0.0024}$ & $0.038 \pm 0.003$ \\
$^{76}$Ge & $1.65^{+0.14}_{-0.12} \cdot10^{21}$ & $0.1122^{+0.0043}_{-0.0045}$ 
& $0.1143^{+0.0044}_{-0.0046}$ & $0.113 \pm 0.006$ \\
$^{82}$Se & $(0.92 \pm 0.07)\cdot10^{20}$ & $0.0826^{+0.0032}_{-0.0031}$ 
& $0.0831^{+0.0033}_{-0.0031}$ & $0.083 \pm 0.004$ \\
$^{96}$Zr & $(2.3 \pm 0.2)\cdot10^{19}$ & $0.0798^{+0.0037}_{-0.0032}$ 
& $0.0804^{+0.0038}_{-0.0033}$ & $0.080 \pm 0.004$ \\
$^{100}$Mo & $(7.1 \pm 0.4)\cdot10^{18}$ & $0.2065^{+0.0061}_{-0.0056}$
& $0.2088^{+0.0062}_{-0.0057}$ \\
  &  & $0.1847^{+0.0050^{a)}}_{-0.0031}$ & & $0.185 \pm 0.005$ \\
$^{100}$Mo- & $6.7^{+0.5}_{-0.4}\cdot10^{20}$ 
& $0.1571^{+0.0048}_{-0.0056}$ &  $0.1619^{+0.0050}_{-0.0058}$  \\
 $^{100}$Ru$(0^{+}_{1})$ &  & $0.1513^{+0.0047^{a)}}_{-0.0053}$ & & $0.151 \pm 0.005$ \\ 
$^{116}$Cd & $(2.87 \pm 0.13)\cdot10^{19}$ & $0.1123^{+0.0026}_{-0.0024}$ 
& $0.1139^{+0.0026}_{-0.0025}$\\
 &  & $0.1049^{+0.0024^{a)}}_{-0.0023}$ & & $0.105 \pm 0.003$\\
$^{128}$Te & $(2.0 \pm 0.3)\cdot10^{24}$ & $0.0431^{+0.0037}_{-0.0029}$ 
& $0.0483^{+0.0042}_{-0.0034}$ & $0.046 \pm 0.006$ \\
$^{130}$Te & $(6.9 \pm 1.3)\cdot10^{20}$ & $0.0308^{+0.0034}_{-0.0026}$
&  $0.0317^{+0.0034}_{-0.0026}$ & $0.031 \pm 0.004$\\
$^{136}$Xe & $(2.19 \pm 0.06)\cdot10^{21}$ & $0.0177^{+0.0003}_{-0.0002}$ 
& $0.0185^{+0.0003}_{-0.0002}$ & $0.0181 \pm 0.0007$ \\
$^{150}$Nd & $(8.2 \pm 0.9)\cdot10^{18}$ & $0.0579^{+0.0034}_{-0.0029}$ 
& $0.0587^{+0.0034}_{-0.0030}$ & $0.058 \pm 0.004$ \\
$^{150}$Nd- & $1.2^{+0.3}_{-0.2}\cdot10^{20}$
& $0.0438^{+0.0042}_{-0.0046}$ & $0.0450^{+0.0043}_{-0.0048}$ & $0.044 \pm 0.005$\\
$^{150}$Sm($0^{+}_{1}$) & & & &\\
$^{238}$U & $(2.0 \pm 0.6)\cdot10^{21}$ & $0.1853^{+0.0361}_{-0.0227}$ 
& $0.0713^{+0.0139}_{-0.0088}$ & $0.13^{+0.09}_{-0.07}$  \\
$^{130}$Ba,  & $ \sim 10^{21}$ & $ \sim 0.26$ \cite{KOT13} & & $ \sim 0.26$ \\
ECEC(2$\nu$) & & & &\\

\hline
\end{tabular}
\end{table}

\subsection{$^{130}$Ba (ECEC)}

For $^{130}$Ba positive results were obtained in geochemical measurements only. 
In geochemical experiments it is not possible to recognize  
the different modes. But I believe that exactly ECEC(2$\nu$) 
process was detected because other modes are strongly suppressed (see, for example, 
estimations in \cite{DOM05,SIN07,BAR13}).
First positive 
result for $^{130}$Ba was mentioned in Ref. \cite{BAR96}, in which experimental 
data from Ref. \cite{SRI76} were analyzed. In this paper 
positive result was obtained for one sample of barite 
($T_{1/2} = 2.1^{+3.0}_{-0.8} \cdot 10^{21}$ yr), but for second sample only 
limit was established ($T_{1/2} > 4 \cdot 10^{21}$ yr). Then more accurate 
half-life values, $(2.2 \pm 0.5) \cdot 10^{21}$ yr \cite{MES01} and 
$(0.60 \pm 0.11)\cdot 10^{21}$ yr \cite{PUJ09}, were obtained. 
However, the results are in 
strong disagreement. One can not use usual 
average procedure in this case. One just can conclude that half-life of $^{130}$Ba is 
$\sim 10^{21}$ yr. 
To obtain more precise and correct half-life value for $^{130}$Ba 
new measurements are needed.

\section{NME values for two neutrino double beta decay}

A summary of the half-life values are presented in Table 2 (2-nd column). 
From the measured half-life one can extract the "experimental" nuclear matrix element 
using the relation \cite{KOT12} 

\begin{equation}
T_{1/2}^{-1} = G_{2\nu}\cdot g_A^4\cdot(m_ec^2\cdot M_{2\nu})^2, 
\end{equation}
 
where $T_{1/2}$ is the half-life value in [yr], $G_{2\nu}$ is the phase space factor in [yr$^{-1}$], 
$g_A$ is the dimensionless axial vector coupling constant 
and  $(m_ec^2\cdot M_{2\nu})$  is the dimensionless nuclear matrix element. 
It is necessary to take into account that there are various indications that 
in nuclear medium the matrix elements 
of the axial-vector operator are reduced in comparison with their free nucleon values. 
This quenching is often described as a reduction of the coupling constant $g_A$ from 
its free nucleon value of $g_A = 1.2701$ \cite{BER12} to the value of 
$g_A \sim 0.35-1.0$ (see discussions in \cite{FAE08,CAU12,SUH13,BAR13a}). 
So, follow the Ref. \cite{KOT12} it is better
to have a deal with so-called "effective" NME, 
$\mid M^{eff}_{2\nu}\mid = g_A^2\cdot \mid (m_ec^2\cdot M_{2\nu})\mid$. And this value has been calculated 
for all mentioned above isotopes.    

The results of these calculations are presented 
in Table 2 (3-d and 4-th columns). To do the calculations I used the $G_{2\nu}$ values from Ref. \cite{KOT12} 
and \cite{STO13,STO14}\footnote{Ref. \cite{STO14} was published as up-date of Ref. \cite{STO13}.
And, finally, I used in this work results of calculations from Ref. \cite{STO14}.}, 
respectively (see Table 3). For $^{130}$Ba $G_{2\nu}$ value 
for ECEC transition was taken from \cite{KOT13}.
These recent calculations pretend to be most reliable and correct
by this moment (see discussions in \cite{KOT12,KOT13,STO13,STO14}). 
Results of these calculations 
are in quite good agreement ($\sim$ 1-7\%) with two exceptions, for $^{128}$Te ($\sim 20\%$) and $^{238}$U.
For $^{238}$U two absolutely different values $14.57\cdot 10^{-21} yr^{-1}$ \cite{KOT12} 
and $98.51\cdot 10^{-21} yr^{-1}$ \cite{STO14})
were obtained. It is clear that calculations for $^{238}$U have to be checked.
For $^{100}$Mo, $^{100}$Mo-$^{100}$Ru$(0^{+}_{1})$ and $^{116}$Cd I used  $G_{2\nu}$ calculated 
in Ref. \cite{KOT12} for SSD 
mechanism, in addition. Corresponding values for $\mid M^{eff}_{2\nu}\mid$ are presented in Table 2 
and
these are most correct values for these isotopes.    
So-called recommended values for $\mid M^{eff}_{2\nu}\mid$ are presented in Table 2 (5-th column)
too. These values were obtained as an average of two values, specified in columns 3 and 4. 
The error of recommended values is chosen so that to cover all range of values from columns 
3 and 4 (taking into account corresponding errors). For $^{100}$Mo, $^{100}$Mo-$^{100}$Ru$(0^{+}_{1})$ 
and $^{116}$Cd I recommend to use values obtained using $G_{2\nu}$ for SSD mechanism.
  
For the majority of isotopes we now have 
$\mid M^{eff}_{2\nu}\mid$ with an accuracy of
$\sim 3-8 \%$. For $^{128}$Te and $^{130}$Te it is $\sim$ 13\% and for $^{150}$Nd-$^{150}$Sm$(0^{+}_{1})$ it is 
$\sim 16\%$. 
The most unsatisfactory situation is for $^{238}$U ($\sim 70\%$) and $^{130}$Ba ($\sim 50\%$).  
For $^{238}$U main uncertainty is connected with accuracy of $G_{2\nu}$ and for $^{130}$Ba with 
accuracy of experimental data for the half-life.

In a few recent publications \cite{CAU12,SUH13,BAR13a} attempts to reproduce NME values 
for two-neutrino double beta decay 
within various models were realized. The conclusion was that 
renormalization (quenching) of $g_A$ is needed to reproduce the experimental data. So within 
Interacting Shell Model (ISM) approach \cite{CAU12} NMEs for $^{48}$Ca, $^{76}$Ge, $^{82}$Se, 
$^{128}$Te, $^{130}$Te and $^{136}$Xe were calculated, using the $g_A$ $\sim 0.57-0.94$ 
(these values were obtained from data for a single beta decay or charge exchange reactions). 
As a result it was succeeded to obtain rather good agreement between calculations 
and experimental data (nevertheless, NME calculated values for $^{82}$Se, $^{128}$Te, 
$^{130}$Te and $^{136}$Xe exceed experimental data on $\sim 20-30\%$). 
In Ref. \cite{SUH13} within QRPA model calculated values of NMEs for $^{100}$Mo, 
$^{116}$Cd and $^{128}$Te were adjusted to experimental values at the expense of 
a choice of the corresponding $g_A$ values ($\sim 0.4-0.75$). The same procedure 
was executed within IBM-2 model for many nuclei \cite{BAR13a}. It was shown that 
for exact reproduction of experimental data the $g_A$ has to be $\sim 0.35-0.71$. 
The question of whether or not the quenching of $g_A$ is the same in $2\nu\beta\beta$ as 
in $0\nu\beta\beta$ decay 
is the subject of debate, but it is clear that this question has to be carefully 
investigated because changes in $g_A$ leads to changes in sensitivity to effective 
Majorana neutrino mass $\langle m_{\nu}\rangle$ in double beta decay experiments.

I would like to note that in all these cases \cite{CAU12,SUH13,BAR13a} when 
comparing with experimental data the 
recommended $T_{1/2}(2\nu)$ values from our previous work \cite{BAR10} were used. 
In the present work more precise experimental values for $T_{1/2}(2\nu)$  and NME 
for many nuclei are obtained and, I believe, that will help with a solution of 
the $g_A$ problem in the future.   

\begin{table}[ht]
\caption{Phase-space factors from Ref.\cite{KOT12}, \cite{STO14} and \cite{KOT13}. 
$^{a)}$ Obtained using SSD model.}
\begin{tabular}{ccc}
\hline
Isotope & $G_{2\nu} (10^{-21} yr^{-1})$ \cite{KOT12} & $G_{2\nu} (10^{-21} yr^{-1})$ \cite{STO14} \\

\hline
$^{48}$Ca & 15550 & 15536 \\ 
$^{76}$Ge & 48.17 & 46.47 \\
$^{82}$Se & 1596 & 1573 \\
$^{96}$Zr & 6816 & 6744 \\
$^{100}$Mo & 3308 & 3231 \\
 & $4134^{a)}$ & \\
$^{100}$Mo-$^{100}$Ru$(0^{+}_{1})$ & 60.55  & 57.08 \\
  & $65.18^{a)}$ & \\
$^{116}$Cd & 2764 & 2688 \\
  & $3176^{a)}$ &  \\
$^{128}$Te & 0.2688 & 0.2149 \\
$^{130}$Te & 1529 & 1442 \\
$^{136}$Xe & 1433 & 1332 \\
$^{150}$Nd & 36430 & 35397 \\
$^{150}$Nd-$^{150}$Sm($0^{+}_{1}$) & 4329 & 4116 \\
$^{238}$U & 14.57 & 98.51 \\
$^{130}$Ba, ECEC(2$\nu$)   & 15000 \cite{KOT13}&   \\

\hline
\end{tabular}
\end{table}

\section{Conclusion}

In summary, all positive $2\nu\beta\beta$-decay results were analyzed, 
and average values for half-lives were calculated. For the cases of 
$^{128}$Te and $^{130}$Ba, the so-called recommended values 
have been proposed. Using these half-life values, $\mid M^{eff}_{2\nu}\mid$  
for two neutrino double beta decay 
were obtained. Finally, previous results from Ref. \cite{BAR10} were successfully up-dated 
and new results for $^{136}$Xe 
were added.
A summary is collected in Table 2. 
I strongly recommend the use of these values as 
the most reliable presently. 

Notice that the accurate half-life values for $2\nu\beta\beta$ decay
could be used to adjust the most relevant parameter of the 
quasiparticle random-phase approximation (QRPA) model, the strength of the 
particle-particle interaction $g_{pp}$. In addition effective $g_A$ value could be established
for 2$\beta$ decay. 
It will give the possibility to improve 
the quality of NME calculations for neutrinoless double beta decay and, finally,
to improve the quality of neutrino mass $\langle m_{\nu} \rangle$ estimations.





\bibliographystyle{elsarticle-num}
\bibliography{<your-bib-database>}




\end{document}